\documentclass[aps,prr,nofootinbib, twocolumn,superscriptaddress,longbibliography]{revtex4-2}

\usepackage[english]{babel}
\usepackage{amsmath,amssymb,bm}
\usepackage{graphicx}
\usepackage{epstopdf}
\usepackage{amsfonts}
\usepackage{amssymb}
\usepackage{amsbsy}
\usepackage{amsmath}
\usepackage{latexsym}
\usepackage{sansmath}
\usepackage{enumitem}
\usepackage{subfigure}
\usepackage{lipsum}
\usepackage{float}
\usepackage{cancel}

\usepackage{bm}
\usepackage{xcolor}
 
\usepackage{comment}
\usepackage{csquotes}								
\usepackage{physics}
 
\usepackage{accents}

\usepackage{hyperref}
\usepackage{nameref}

\newcommand{\cl}{\Lambda}

\newcommand{\be}{\begin{eqnarray}}
\newcommand{\ee}{\end{eqnarray}}
\def\p{\partial}
 \def\nn {\nonumber}

\begin{document}

\title{Shock waves in classical dust collapse}

\author{Viqar Husain} \email{vhusain@unb.ca}
\affiliation{Department of Mathematics and Statistics, University of New Brunswick, Fredericton, Canada.}

\author{Hassan Mehmood} \email{hassan.mehmood@unb.ca}
\affiliation{Department of Mathematics and Statistics, University of New Brunswick, Fredericton, Canada.}

 \date{\today}

\begin{abstract}

During gravitational collapse of dust in spherical symmetry, matter particles may collide forming shell crossing singularities (SCS) at which the Einstein equations become indeterminate. We show that in the case of marginally bound dust collapse, there is a unique evolution beyond SCS such that a propagating shock wave forms, the metric remains continuous, and the stress-energy tensor dynamically becomes that of a thin shell. We give numerical simulations that exhibit this result.  

\end{abstract}

\maketitle


One of the most interesting problems in classical general relativity is the dynamics of gravitational collapse. The simplest setting for studying this problem is in spherical symmetry because the long-time static limit of gravitational collapse is expected to be the Schwarzschild black hole, the first exact solution of Einstein's equation.  This solution is the universal attractor provided matter satisfies physically reasonable conditions, such as positive energy density and timelike or null matter flux. 

How matter behaves as it undergoes gravitational collapse is also an important question. The details depend on the type of matter, and whether or not it has internal pressure. There are two well-studied exact solutions where the answer is known to some extent: for null fluid, the solution is the Vaidya metric \cite{Vaidya} and its generalization to include pressure \cite{Husain_VaidyaGen}; for timelike dust, the solution is  the Lemaitre-Tolman-Bondi (LTB) spacetimes \cite{Bondi:1947fta, Tolman:1934za}. In the former case, particles of the fluid follow infalling lightlike trajectories until all the fluid's mass-energy enters its Schwarzschild radius and a black hole forms; in the latter case there is the additional interesting possibility that timelike dust trajectories intersect at a location outside the Schwarzschild radius, a situation known as ``shell crossing." 

For a brief review of the problem let us begin with the spherically symmetric metric 
\be
ds^2 = -dt^2 + X(R,t)^2 dR^2 +r^2(R,t)d\Omega^2. \label{LTBm}
\ee
Then the Einstein equations with pressureless dust $G_{ab} = 8\pi \rho\ u_a u_b$ with $u^a = (\p/\p t)^a$ lead to \cite{Bondi:1947fta}
\be
X &=& \frac{r'^2}{1+E(R)} \label{X}\\
E(R) &=&\dot{r}^2 - \frac{2M(R)}{r} \label{E}
\ee
where $\dot{r} = \p r/\p t,\ r' = \p r/\p R$, $M(R)$ and $E(R)$ are integration functions, and the dust density is 
\be
\rho &=& \frac{M'}{4\pi r^2 r'}.\label{rho}
\ee
This is the LTB solution. It describes non-interacting particles (dust ``shells") at each value of $R$ with ``energy" $E(R)$. The metric (in these coordinates) is degenerate where $r'(R,t)=0$ according to (\ref{X}), and the density (\ref{rho}) is divergent provided $M'\ne 0$ at the radial points in question. Also at such points, the proper radial distance $ds^2_R = X^2(R,t)\ dR^2$ between the shells labeled by $R$ vanishes, hence the term ``shell crossing singularity" (SCS). The second type of metric degeneracy occurs where $r(R,t)=0$, when a shell reaches the center of collapse. Our focus will be on SCS with the assumption that $E(R)= 0$.   

There is a significant literature on the nature of SCS. It is known that at an SCS the curvature singularity can be naked \cite{yodzis1973occurrence}, and is gravitationally weak in the sense that tidal forces  \cite{Newman:1985gt,Nolan:1999tw} and redshifts are bounded \cite{Hellaby:1985zz}. It has been shown that if initial data are suitably restricted that SCS do not occur \cite{lake-84}. However, this ``cure" appears too restrictive from the physical perspective of permitting all regular initial data.

There have also been attempts to extend the LTB spacetime beyond an SCS: the first such attempt was for a specific class of initial data \cite{Papapetrou:1973gy}, followed by a more general proposal for extension \cite{clarke-donnell}; it was shown in \cite{Frauendiener} that additional information about shell interaction is required for spacetime extension. Another approach used the fact that the LTB equations may be written as a conservation law that may be used to find solutions of the integrated version of the equations, called weak solutions. Such solutions are known to lead to shock waves in fluid mechanical systems. An issue is that weak solutions are not unique, they depend on the choice of variables used. For the LTB system this approach yields a shock wave such that the metric is bounded but discontinuous at the shock \cite{Nolan:2003wp, Nolan:2004dm}. A more recent method suggests excising the SCS, and replacing it with a thin shell \cite{booth} using Israel junction conditions \cite{Israel:1966rt, Israel:1967zz, Poisson:2009pwt, papapetrou-68}; this yields a continuous metric throughout spacetime.  

As may be gathered by this summary highlighting the main approaches to extending LTB metrics beyond SCS, there is no consensus on what is the unique physics. The approach that uses weak solutions appears to be the most natural as it is well established in fluid mechanics and similar systems; the issue of their non-uniqueness can be settled by deriving conservation laws from symmetries, such as the conservation laws for rest mass, energy and momentum in Newtonian physics.

Here we show that in the case of marginally bound dust collapse, (i) there are weak solutions of the LTB system in which a shock wave forms at SCS such that the metric is continuous\footnote{After this work was submitted for publication, a  related result for the quantum corrected LTB model appeared on the arXiv \cite{Liu:2025fil}} at the shock and the dynamics extends naturally beyond SCS, (ii) the variable for which this occurs is unique, and (iii) the dynamical equation for the shock is different from that of the thin shell derived using the junction conditions. These results settle the long-standing question of extending the LTB metric beyond SCS. We also present numerical examples of evolution that lead to the formation of shock waves using techniques applicable to hyperbolic conservation laws \cite{lev02, leveque-92}. To arrive at these results, we formulate the gravity-dust Einstein equations in the canonical theory in the dust time gauge \cite{Lasky:2006hq,Husain:2011tk}.  

Our starting point for deriving the canonical gravity-dust equations begins with the spatial metric in the form
\be 
ds^2 = \Lambda^2 dr^2 + Y^2 d\Omega^2,
\ee
and the 3+1 form of spherically symmetric action \cite{Kuchar:1994zk} 
\be 
    S_{EH} &=& \int dt\int dr\,[P_\Lambda\Dot{\Lambda} + P_Y \dot{Y} - NH - N^rH_r], \\
    S_D &=& \int dt\int dr [p_T\Dot{T} - NH_D - N^rC_D].
\ee
 where $P_\Lambda$ and $P_Y$ are the  momenta conjugate to the metric variables $\Lambda$ and $Y$, $H$ and $H_r$ are the Hamiltonian and spatial diffeomorphism constraints, respectively. Similarly, $p_T$ is the momentum conjugate to the dust field $T$, and $H_D$ and $C_D$ are the dust Hamiltonian and spatial diffeomorphism constraints. Explicitly, 
\be 
    H &=& -\frac{P_YP_\cl}{Y} +  \frac{\cl P^2_\cl}{2Y^2} + \frac{YY''}{\cl} - \frac{YY'\cl'}{\cl^2} \nn\\
     && +  \frac{(Y')^2}{2\cl} - \frac{\cl}{2}, \label{hami} \\
    H_r &=& P_YY' - \cl P'_\cl,\ \nn\\
  H_D &=& \sqrt{p_T^2 + \frac{(p_TT')^2}{\cl^2}},  \nn\\
  C_D &=& -p_TT', \label{diffeo}
\ee
where the prime denotes differentiation w.r.t $r$. The dust time gauge $T=t$ leads to lapse $N=1$ \cite{Husain:2011tk}, and the areal gauge $Y=r$ fixes the radial shift  to be 
\be
N^r = P_\Lambda/r. \label{Nr}
\ee
The final result is the physical Hamiltonian
\be
    H_{\text{phy}} = \int_0^\infty dr\, \left[\frac{1}{2\cl} - \frac{\cl}{2}  -\frac{\cl P_\cl P'_\cl}{r} + \frac{\cl P^2_\cl}{2r^2} - \frac{r\cl'}{\cl^2}  \right];
\ee
This leads to the evolution equations
\be
    \dot{P}_\cl &=& \{P_\cl, H_{\text{phy}}\} = \left( \frac{P_\cl^2}{2r} \right)^\prime - \frac{1}{2}\left(\frac{1}{\cl^2}-1 \right) , \label{eoms-pl}\\ 
    \dot{\cl} &=& \{\cl, H_{\text{phy}}\} = \frac{\cl'P_\cl}{r}. \label{eoms-lam}
\ee
The dust energy density in these variables is given by \cite{Husain:2011tk}
\be
 \rho(t,r) = -\frac{H_{\text{phys}}}{4\pi\cl r^2} \label{rho2}
\ee
For comparison with the functions $M$ and $E$ in eqns. (\ref{X}-\ref{E}), we note the following relations in transforming from the $(t,R)$ coordinates used in (\ref{LTBm}) to the coordinates $(t,r)$ in which (\ref{eoms-pl}--\ref{eoms-lam}) are written \cite{Lasky:2006hq}: 
\be
\Lambda^2 = \left(1+E \right)^{-1},\quad \frac{P_\Lambda}{r} = \sqrt{ E+\frac{2M}{r}}.
\ee

The so-called marginally bound LTB solutions, defined to be those with $E(R)=0$ in (\ref{X}), correspond to $\Lambda =1$ \cite{Lasky:2006hq}. This is the case we consider to show our main result. With $\Lambda =1$, the remaining dynamical equation (\ref{eoms-pl}) becomes the conservation law
\be
\dot{P}_\cl - \left( \frac{P_\cl^2}{2r} \right)^\prime=0. \label{PLeqn}
\ee
The second term in this equation has an appealing interpretation if written in terms of the Misner-Sharp-Wheeler (MSW) mass function \cite{misner-sharp-mass}. Noting that the radial shift function (\ref{Nr}) may be rewritten in terms of the MSW mass function as 
\be 
N^r(r,t) = \frac{P_\Lambda(r,t)}{r} = \sqrt{\frac{2M(r,t)}{r}};
\label{MP}
\ee
eqn. (\ref{PLeqn}) becomes 
\be
\partial_t{P}_\cl-\partial_rM(r, P_\cl)=0,
\label{PMeqn}
\ee
and the dust density (\ref{rho2}) simplifies to 
$\rho = M'/4\pi r^2$ confirming the interpretation of $M(r,t)$ as the MSW mass function.  

It is possible to write (\ref{PLeqn}) entirely in terms of $M$ as  
\be 
\p_t\left(\sqrt{r}M\right) -\p_r\left[\frac{(2M)^{3/2}}{3} \right] =0,\label{Meqn}
\ee
Thus the evolution equation for the marginally bound LTB case may be written as conservation laws in (at least) two different ways, (\ref{PLeqn}) and (\ref{Meqn}). More generally, a conservation law can be written in terms of an arbitrary function of $M$ \cite{Nolan:2003wp}. As we shall see, different ways of writing down the conservation law may lead to different extensions of the spacetime beyond SCS; we will argue that of these infinitely many laws, \eqref{PLeqn} should be preferred, for it gives rise to a metric that is continuous across SCS. For $E=0$, the metric takes the PG form
 \be  
ds^2 = -dt^2 + (N^rdt +dr)^2 + r^2 d\Omega^2 \label{PGm}
 \ee
 
We now proceed to study the two equations (\ref{PLeqn}) and (\ref{Meqn}), and show that they give rise to shock waves with substantially different properties of the metric, specifically that the solution to (\ref{PMeqn}) provides a continuous extension of the metric beyond the SCS, while that of (\ref{Meqn}) does not. This is our main result. 

The characteristic equations of (\ref{PMeqn}) giving the flow on the solution surface $(r(s),t(s), P_\Lambda(s))$ are
\be 
\frac{dP_\Lambda}{ds} = -\frac{P_\Lambda^2}{2r^2}; \quad \frac{dr}{ds} = -\frac{P_\Lambda}{r}; \quad \frac{dt}{ds} =1. 
\label{char123}
\ee
These equations are equivalent to the equations of motion \eqref{E} in LTB coordinates, with $E=0$. SCS in this context correspond to characteristic crossings. At a characteristic crossing, shells of different mass (and hence different $P_\cl$) collide at the same radius; thus the speed $dr/ds$ in \eqref{char123} becomes multi-valued and the method of characteristics breaks down. Thus, some other method must be employed to continue a solution beyond SCS.

\begin{figure}[hbtp]
   \includegraphics[width=\columnwidth, height=6in]{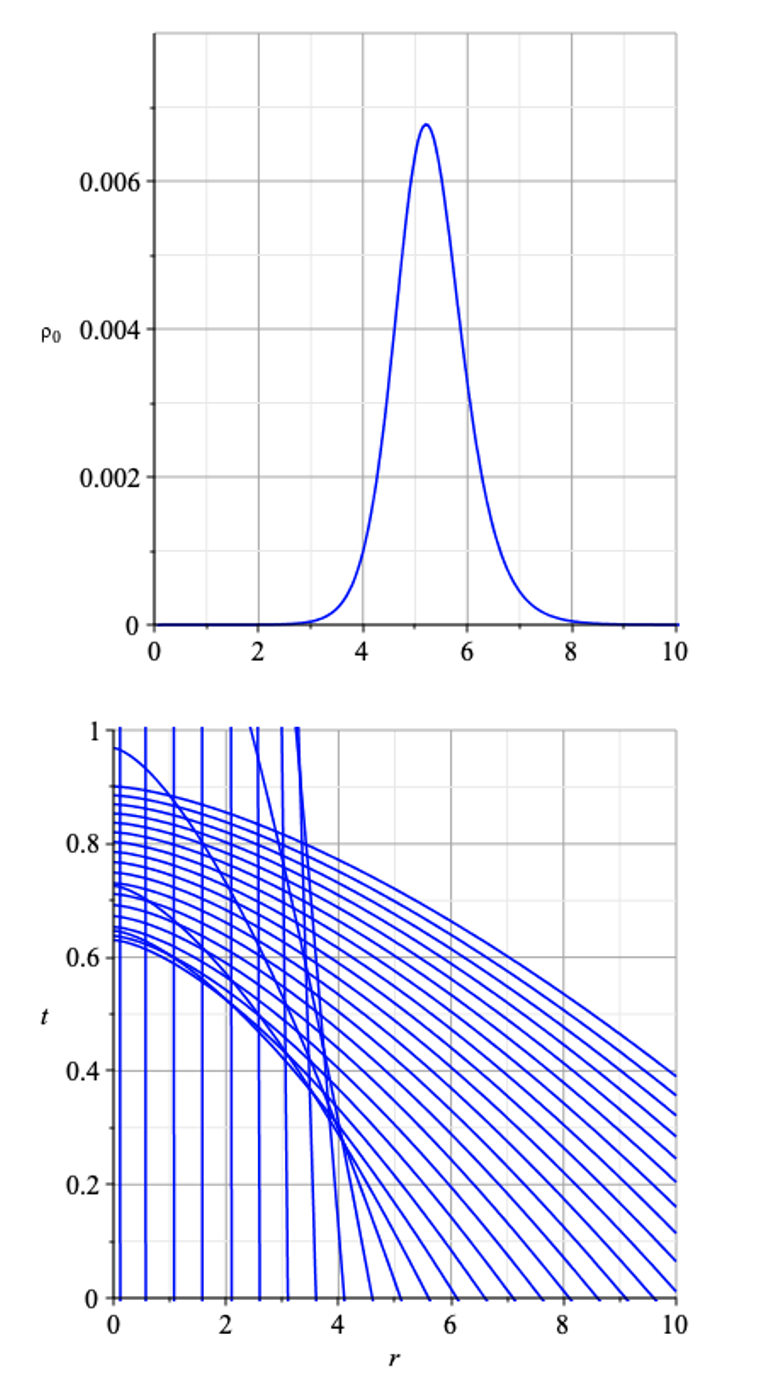}
    \caption{ An example of characteristic crossings for the initial density in the upper frame. The region to the left of the  matter distribution is Minkowski spacetime where the characteristics are vertical lines.}
    \label{charplot}
\end{figure}

It is important emphasize that characteristic crossings  occurs generically for asymptotically flat initial data as shown in Fig. \ref{charplot} : the curves to the left of the mass distribution are vertical as this region is close to Minkowski spacetime, and the curves to the right curve toward $r=0$ due to the mass within. (See also \cite{Fazzini:2023ova} for a result for effective LTB models.) 

\smallskip
\noindent\underbar{\bf Weak solution} Continuing a solution beyond the SCS requires consideration of {\it weak solutions}, which are solutions of the integrated version of eqn.(\ref{PMeqn}). For radial integration domain $[r_1,r_2]$ this is 
\be 
&&\frac{d}{dt} \int_{r_1}^{r_2} P_\Lambda (r,t)\ dr \nn\\
&&=M(r_2, P_\Lambda(t, r_2)) - M(r_1, P_\Lambda(t, r_1)),
\label{weakPM}
\ee
i.e. the change of $P_\Lambda$ in the radial interval is due to the net flux of $M$ through its boundaries. It is worth emphasizing how the integrated version of the equation of motion allows one to continue solutions beyond SCS. As noted above, at a SCS, the solution becomes discontinuous. Hence, it ceases to be a solution of the \textit{differential} equation. However, it is nonetheless possible to \textit{integrate} a large class of discontinuous functions (and their derivatives). In this way solutions to the integral equation tell us how the discontinuities arising as a result of SCS propagate beyond the time of their first appearance; such propagating discontinuities are called \textit{shock waves} in the literature on hyperbolic conservation laws \cite{leveque-92, Dafermos:1315649}. 

To see how this all works in practice, let us consider a specific (pedagogical) example of a weak solution: a thin shell of dust such that $P_\Lambda$ are constants $P_\cl^L,P_\cl^R$ on either side of the thin shell. This could be the situation where all fluid elements have collided at a particular radius $r=S$, forming a thin shell with a delta function in the energy-momentum tensor. Then, at time $t=0$ the initial data is 
\be 
P_\cl(0,r) = \Theta(r-S)P^R_\cl + \Theta(S-r)P^L_\cl, \label{riemann-data}
\ee
Finding solutions of such data is known as the Riemann initial-value problem. We expect dynamics to be such that the surface of discontinuity is propagating, i.e. $S= S(t)$, therefore we can write
\begin{equation}
    P_\cl(t,r) = \Theta(r-S(t))P^R_\cl + \Theta(S(t)-r)P^L_\cl. \label{riemann-sol}
\end{equation}
It remains to find the shock speed $dS(t)/dt$. This is done by substituting (\ref{riemann-sol}) into (\ref{weakPM}). With $r_1 = S(t)-\delta r$ and $r_2 = S(t) + \delta r$ in \eqref{weakPM}, and recalling that $\Dot{\Theta}(g(t))=\Dot{g}(t)\delta(g(t))$, one gets 
\be
    \frac{dS(t)}{dt} &=& -\frac{M(S(t), P^R_\cl)-M(S(t),P^L_\cl)}{P^R_\cl-P^L_\cl} \label{rh-condition}\\ 
    &=& -\frac{P^R_\cl + P^L_\cl}{2S(t)}, \label{shock-speed-1}
\ee
where the second equality follows from (\ref{MP}), namely $M=P^2_\cl/2r$. Eqn. \eqref{rh-condition} is the so-called Rankine-Hugoniot (RH) jump condition for this system. It is revealing to compare the last equality with the second characteristic equation (\ref{char123}) --- the r.h.s of (\ref{shock-speed-1}) is just the average of $P_\cl$ across the shell. 

Although the above illustrative derivation of the RH condition is restricted to the case when the solution on either side of a SCS is constant, the result is much more general. For generic initial data, instantaneous SCS arise such that on either side of them, the solution is a time-dependent solution of the equation of motion; by flux matching, as done above, the shock wave speed is of the same form \eqref{shock-speed-1} with $P^R_\cl$ and $P^L_\cl$ now time-dependent (see e.g., \cite{leveque-92}). This is what is captured in the numerical method described below.


\smallskip

\noindent\underbar{\bf Metric continuity} We now ask what is the implication of the shock velocity (\ref{shock-speed-1}) for the metric (\ref{PGm}), i.e. is the metric continuous at the shock? This is answered by computing the induced metric at the shock surface $\Sigma$ from both sides by substituting 
\be
dr|_{\Sigma} = dS(t) = -\frac{P^R_\cl(t) + P^L_\cl(t)}{2S(t)}dt
\ee
into (\ref{PGm}). This gives
\be 
    ds^2|_{\Sigma_R} &=&  -\left[1 -\left(\frac{P^R_\cl -P^L_\cl}{2S} \right)^2\right]dt^2 + S^2d\Omega^2 \nn\\
    &=& ds^2|_{\Sigma_L}.  \label{shock-metric}
\ee
Thus the metric is continuous at the shock for the dynamical equation (\ref{PMeqn}). It is important to emphasize that this continuity property is {\it derived} from the weak solution, unlike the case of Israel junction conditions where it is imposed.  
\smallskip

\noindent\underbar{\bf Uniqueness} We now show a uniqueness property: There is no shock speed other than (\ref{shock-speed-1}) that leads to continuity of the metric at the shock. This question is useful to address because the dynamical equation may be written as a conservation law in more than one variable, as noted above in 
eqns. (\ref{PMeqn}) and (\ref{Meqn}). Assume that the shock speed is the arbitrary function
\be
\frac{dS(t)}{dt} = F(S,P_\cl^R,P_\cl^L)
\ee
and consider the metric of the form (\ref{PGm}). Then the continuity of the metric at the shock requires
\be
 \left(\frac{P_\cl}{S}  + F \right)^2_{|L}= \left(\frac{P_\cl}{S}  + F \right)^2_{|R}.
\ee
The unique solution for $F$ is the r.h.s. of (\ref{shock-speed-1}). This is one of our main results. 

As an example, one can compute the shock speed that arises from the alternative dynamical equation (\ref{Meqn}): the result is 
\be
    \frac{dS(t)}{dt} &=& -\frac{2\sqrt{2}}{3\sqrt{S(t)}}\frac{(M^R)^{3/2}-(M^L)^{3/2}}{M^R-M^L} \nn\\
    &=& -\frac{2}{3S(t)}\frac{(P^R_\cl)^3-(P^L_\cl)^3}{(P^R_\cl)^2-(P^L_\cl)^2}, 
    \label{shock-speed-3}
\ee
By a calculation similar to the above, it follows that the metric is not continuous at the shock. 

It may seem paradoxical that a mere change of variables in a differential equation (cf. eqn. \eqref{PLeqn} and \eqref{Meqn}) can lead to distinct weak solutions, eqn. \eqref{shock-speed-1} and \eqref{shock-speed-3}. However, the apparent paradox is easily resolved. As we have seen above, finding weak solutions involves integrating discontinuous functions and their derivatives; thus, strictly speaking, weak solutions are supported on the space of distributions on $[0,\infty)\times [0,\infty)$. Now, it is well-known that a product of distributions is an ill-defined operation. Going between eqn. \eqref{PLeqn} and eqn, \eqref{Meqn} involves changing variables $M = P_\cl^2/2r $ and writing down such terms as $P_{\cl}P_{\cl}'$ or $P_{\cl}\dot{P}_{\cl}$, etc.; these operations involve products of distributions and hence have only formal significance. In other words, the equivalence of eqn. \eqref{PLeqn} and \eqref{Meqn} holds only if the functions entering them are smooth, whereas shock-wave solutions do not have this property. (A detailed discussion of these issues appears in \cite{colombeau-92}.)
\smallskip

\noindent\underbar{\bf Thin shell} An example of the shock wave speed (\ref{shock-speed-1}) comes from applying it to a thin shell in vacuum. This well-known model is normally studied using junction conditions \cite{Israel:1966rt, Israel:1967zz, Poisson:2009pwt, papapetrou-68}, but as we now show, the weak solution gives a different dynamics. Consider the shell density 
\begin{equation}
    \rho(t,r) = \frac{m}{4\pi r^2}\ \delta(r-S(t)),
    \label{shell}
\end{equation}
with mass parameter $m$, with the goal of determining $S(t)$ from the weak equation. From this density it follows that the mass function $M(r,t)$ and $P_\cl(r,t)$ are  
\be
    M(t,r) &=& m\ \Theta(r-S(t)),\\
    P_{\cl}(t,r) &=& \sqrt{2mr}\ \Theta(r-S(t)), \label{p-thin-shell}
\ee
where the last equation follows from (\ref{MP}). Since the interior of the thin shell is Minkowski spacetime, $P_\cl^L=0$ in (\ref{shock-speed-1}), hence 

\begin{equation}
    \frac{dS(t)}{dt} = -\sqrt{\frac{m}{2S(t)}}, \label{eom-thin-shell}
\end{equation}
which has the solution
\begin{equation}
    S(t) = \left[S(t_0)^{3/2}-\frac{3}{2}\sqrt{\frac{m}{2}}(t-t_0)   \right]^{2/3}
\end{equation}
where $S(t_0)$  is the initial position of the shell. This is the shell dynamics of the weak solution of (\ref{PMeqn}) with the metric (\ref{PGm}). 
\smallskip

\noindent\underbar{\bf Comparison with junction conditions} Let us now compare the shock equation (\ref{eom-thin-shell})  with that of the thin shell  obtained from the junction conditions. The usual treatment of the thin shell is in Schwarzschild coordinates with the following result \cite{Poisson:2009pwt, Israel:1967zz} for the marginally bound case (where the shell velocity is zero at spatial infinity):

\begin{equation}
    \left(\frac{dR(\tau)}{d\tau}\right)^2 = \left(1 + \frac{m}{2R(\tau)} \right)^2-1, \label{thin-shell-israel}
\end{equation}
where $R(\tau)$ is the radius of the shell and $\tau$ is the shell's proper time for the induced metric on the shell, $ds^2|_{\text{shell}}= d\tau^2 + R(\tau)^2d\Omega^2$. To compare with the shock equation (\ref{eom-thin-shell}), we transform (\ref{eom-thin-shell}) from PG time $t$ to the proper time $\tau$ using  $[1-m/2S(t)]dt^2 = d\tau^2$. This gives 
\be
    \left(\frac{dS(\tau)}{d\tau}\right)^2 = \left(\frac{2S(\tau)}{m} -1\right)^{-1}. \label{shock-tau}
\ee
Comparison of the last two equations shows that the weak solution and the junction conditions give significantly different dynamics, although the metric is continuous at the shock in the first case and at the shell in the second case. To understand the difference, let us note that the ADM momentum in the chosen gauge is \cite{Husain:2005gx}
\be 
\tilde{\pi}^{ab} = \frac{P_\cl}{2}\ r^ar^b + \frac{rP_\cl'}{4}\ (e^{ab}-r^ar^b),
\label{ADM-p}
\ee
where $r^a$ is the unit radial vector and $e^{ab}$ is the Euclidean 3-metric. Now, the second junction condition (the discontinuity in the extrinsic curvature on the timelike shell surface), when projected on the PG spatial slice, gives a discontinuity in the radial component of the ADM momentum, and continuity in the angular directions \cite{Poisson:2009pwt}. According to (\ref{ADM-p}) this translates to a discontinuity in $P_\Lambda$  {\it and} continuity in $P_\Lambda'$. In contrast, the weak solution requires only the former, with no condition on $P_\Lambda'$. This is the reason why weak dynamics is different from junction condition dynamics.

\smallskip

\begin{figure}[hbtp]
   \includegraphics 
   [width=\columnwidth,height=8.0in]{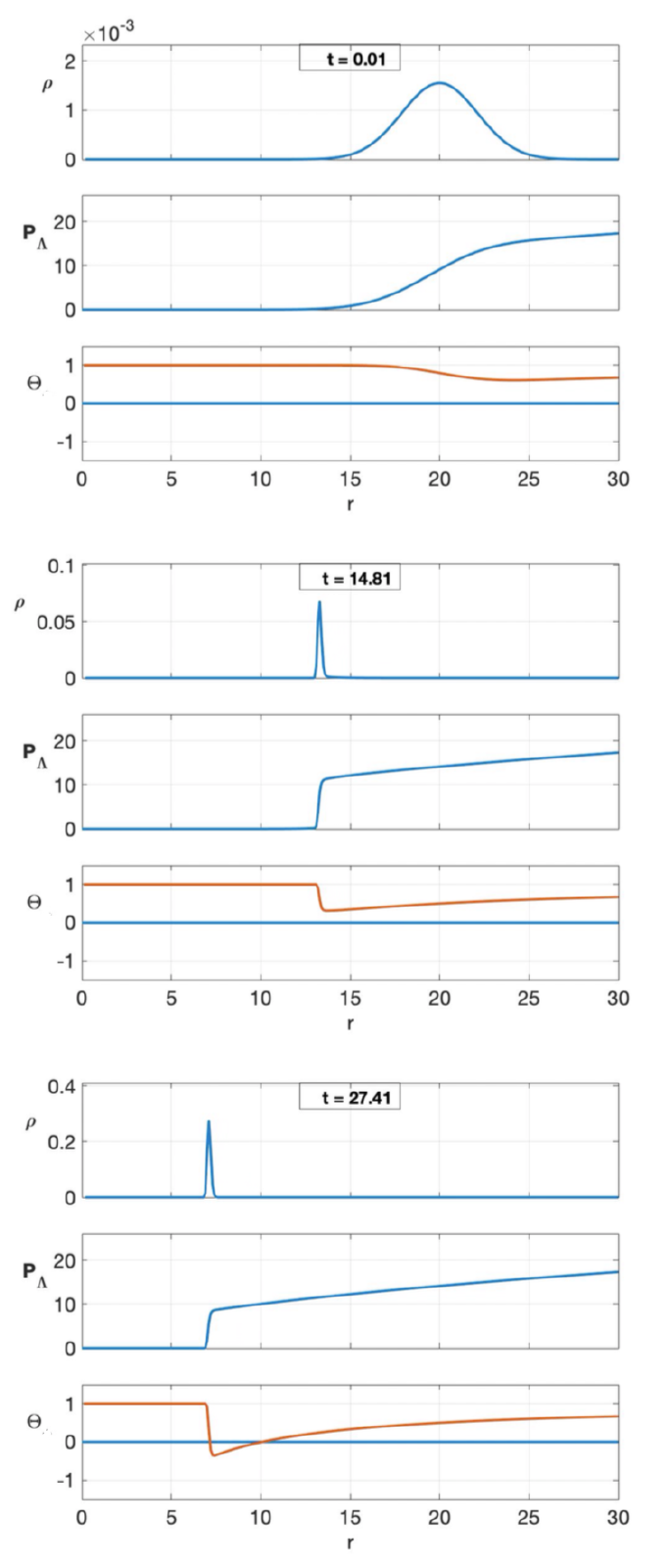}
    \caption{Shock formation: the top frame shows the initial dust density $\rho$, $P_\Lambda$, and apparent horizon function $\Theta$; the middle frame ($t=14.81$) shows shock formation with discontinuity in $P_\Lambda$; the bottom frame ($t=27.41$) shows the shock after entering the event horizon (the outer root of $\Theta$ at $r=10$), with shock located at the inner horizon $r\approx 7.5$.}
    \label{shockplot}
\end{figure}

\noindent\underbar{\bf Numerical solution} The conservation equation (\ref{PMeqn}) can be solved for the weak solution using the standard technique known as the Godunov method \cite{leveque-92}; it was first used for the quantum-corrected LTB model in \cite{Husain:2021ojz, Husain:2022gwp}. This method breaks the spatial integration domain into cells and computes the net flux of $M$ in (\ref{PMeqn}) through each cell at every time step to arrive at a numerical solution. The following numerical simulations demonstrate the analysis of eqn. (\ref{PMeqn}) presented above. 

Fig. \ref{shockplot} shows frames from a typical evolution starting from an initial Gaussian dust profile of ADM mass $5$ (in arbitrary units), and the evolution of $\Theta\equiv|\nabla_a r|^2 = 1- (N^r)^2$, whose roots give the (dynamical) locations of apparent horizons. The dust density profile shrinks in width to form a shock wave near $t=15$, which then falls into the Schwarzschild radius (the outer root of $\Theta$); an inner horizon forms after the first root of $\Theta$ and moves with the shock while the outer horizon remains fixed at twice the radius of the ADM mass at $r=10$. (We note that, unlike the case shown here, there are initial data for which the shock forms inside the Schwarzschild radius of the associated data.) 
 
\smallskip

\noindent\underbar{\bf Discussion} We have presented three results that provide new insight into the classical problem of dust collapse initiated by the LTB solution. These are: (i) a unique extension of the metric beyond SCS such that a shock wave forms with the metric continuous at the shock; this comes from weak solutions of eqn. (\ref{PMeqn}); no other form of the equation gives metric continuity at the shock; (ii) the weak dynamics provided by (\ref{PMeqn}) is different from the junction conditions where metric continuity is imposed rather than derived from the shock speed; (iii) numerical simulations using the Godunov method exhibit our analytical results showing how a smooth initial density dynamically forms a shock wave accompanied by a discontinuity in the variable $P_\Lambda$. 

These results show that there is no reason to exclude initial data that lead to SCS outside the horizon as advocated in \cite{lake-84}, since the formation of shock waves with continuity of the metric beyond SCS follows from the weak solutions of (\ref{PMeqn}). Our results also suggest that there is no reason to propose ad hoc extensions beyond SCS. We considered only marginally bound solutions; the $E\ne 0$ case (\ref{E}) requires an analysis of the coupled equations (\ref{eoms-pl}-\ref{eoms-lam}). This is a more challenging task because neither equation is in the conservation law form in the chosen variables.


\smallskip
\noindent \underbar{Acknowledgements} This work was supported by the Natural Science and Engineering Research Council of Canada. We thank Francesco Fazzini and  Edward Wilson-Ewing for discussions.

\bibliography{refs.bib}

@article{Husain:2021ojz,
    author = "Husain, Viqar and Kelly, Jarod George and Santacruz, Robert and Wilson-Ewing, Edward",
    title = "{Quantum Gravity of Dust Collapse: Shock Waves from Black Holes}",
    eprint = "2109.08667",
    archivePrefix = "arXiv",
    primaryClass = "gr-qc",
    doi = "10.1103/PhysRevLett.128.121301",
    journal = "Phys. Rev. Lett.",
    volume = "128",
    number = "12",
    pages = "121301",
    year = "2022"
}

@article{Husain:2005gx,
    author = "Husain, Viqar and Winkler, Oliver",
    title = "{Flat slice Hamiltonian formalism for dynamical black holes}",
    eprint = "gr-qc/0503031",
    archivePrefix = "arXiv",
    doi = "10.1103/PhysRevD.71.104001",
    journal = "Phys. Rev. D",
    volume = "71",
    pages = "104001",
    year = "2005"
}

@article{misner-sharp-mass,
  title = {Relativistic Equations for Adiabatic, Spherically Symmetric Gravitational Collapse},
  author = {Misner, Charles W. and Sharp, David H.},
  journal = {Phys. Rev.},
  volume = {136},
  issue = {2B},
  pages = {B571--B576},
  numpages = {0},
  year = {1964},
  month = {Oct},
  publisher = {American Physical Society},
  doi = {10.1103/PhysRev.136.B571},
  url = {https://link.aps.org/doi/10.1103/PhysRev.136.B571}
}

@article{Vaidya,
  title = {Nonstatic Solutions of Einstein's Field Equations for Spheres of Fluids Radiating Energy},
  author = {Vaidya, P. C.},
  journal = {Phys. Rev.},
  volume = {83},
  issue = {1},
  pages = {10--17},
  numpages = {0},
  year = {1951},
  month = {Jul},
  publisher = {American Physical Society},
  doi = {10.1103/PhysRev.83.10},
  url = {https://link.aps.org/doi/10.1103/PhysRev.83.10}
}

@article{Husain_VaidyaGen,
  title = {Exact solutions for null fluid collapse},
  author = {Husain, Viqar},
  journal = {Phys. Rev. D},
  volume = {53},
  issue = {4},
  pages = {R1759--R1762},
  numpages = {0},
  year = {1996},
  month = {Feb},
  publisher = {American Physical Society},
  doi = {10.1103/PhysRevD.53.R1759},
  url = {https://link.aps.org/doi/10.1103/PhysRevD.53.R1759}
}

@article{yodzis1973occurrence,
  title={On the occurrence of naked singularities in general relativity},
  author={Yodzis, P and Seifert, HJ and Zum Hagen, H M{\"u}ller},
  journal={Commun. Math. Phys},
  volume={34},
  number={2},
  pages={135--148},
  year={1973}
}

@article{Kuchar:1994zk,
    author = "Kuchar, Karel V.",
    title = "{Geometrodynamics of Schwarzschild black holes}",
    eprint = "gr-qc/9403003",
    archivePrefix = "arXiv",
    reportNumber = "UU-REL-94-3-1",
    doi = "10.1103/PhysRevD.50.3961",
    journal = "Phys. Rev. D",
    volume = "50",
    pages = "3961--3981",
    year = "1994"
}

@article{Husain:2011tk,
    author = "Husain, Viqar and Pawlowski, Tomasz",
    title = "{Time and a physical Hamiltonian for quantum gravity}",
    eprint = "1108.1145",
    archivePrefix = "arXiv",
    primaryClass = "gr-qc",
    doi = "10.1103/PhysRevLett.108.141301",
    journal = "Phys. Rev. Lett.",
    volume = "108",
    pages = "141301",
    year = "2012"
}

@article{Lasky:2006hq,
    author = "Lasky, Paul D. and Lun, Anthony W. C. and Burston, Raymond B.",
    title = "{Initial value formalism for dust collapse}",
    eprint = "gr-qc/0606003",
   journal = "ANZIAM",
   doi = "10.21914/anziamj.v49i0.205",
    archivePrefix = "arXiv",
    volume = "106",
    number = "2",
    year = "2006"
}

@article{Husain:2022gwp,
    author = "Husain, Viqar and Kelly, Jarod George and Santacruz, Robert and Wilson-Ewing, Edward",
    title = "{Fate of quantum black holes}",
    eprint = "2203.04238",
    archivePrefix = "arXiv",
    primaryClass = "gr-qc",
    doi = "10.1103/PhysRevD.106.024014",
    journal = "Phys. Rev. D",
    volume = "106",
    number = "2",
    pages = "024014",
    year = "2022"
}

@article{Hellaby:1985zz,
    author = "Hellaby, C. and Lake, K.",
    title = "{Shell crossings and the Tolman model}",
    doi = "10.1086/162995",
    journal = "Astrophys. J.",
    volume = "290",
    pages = "381",
    year = "1985"
}

@article{Nolan:2003wp,
    author = "Nolan, Brien C.",
    title = "{Dynamical extensions for shell crossing singularities}",
    eprint = "gr-qc/0301028",
    archivePrefix = "arXiv",
    doi = "10.1088/0264-9381/20/4/302",
    journal = "Class. Quant. Grav.",
    volume = "20",
    pages = "575--586",
    year = "2003"
}

@article{Fazzini:2023ova,
    author = "Fazzini, Francesco and Husain, Viqar and Wilson-Ewing, Edward",
    title = "{Shell-crossings and shock formation during gravitational collapse in effective loop quantum gravity}",
    eprint = "2312.02032",
    archivePrefix = "arXiv",
    primaryClass = "gr-qc",
    doi = "10.1103/PhysRevD.109.084052",
    journal = "Phys. Rev. D",
    volume = "109",
    number = "8",
    pages = "084052",
    year = "2024"
}

@article{Israel:1966rt,
    author = "Israel, W.",
    title = "{Singular hypersurfaces and thin shells in general relativity}",
    doi = "10.1007/BF02710419",
    journal = "Nuovo Cim. B",
    volume = "44S10",
    pages = "1",
    year = "1966",
    note = "[Erratum: Nuovo Cim.B 48, 463 (1967)]"
}

@article{Newman:1985gt,
    author = "Newman, Richard P. A. C.",
    title = "{Strengths of naked singularities in Tolman-Bondi space-times}",
    doi = "10.1088/0264-9381/3/4/007",
    journal = "Class. Quant. Grav.",
    volume = "3",
    pages = "527--539",
    year = "1986"
}

@article{Papapetrou:1973gy,
    author = "Papapetrou, A.",
    title = "{Shock waves in the newman-penrose formalism}",
    doi = "10.1007/BF01645682",
    journal = "Commun. Math. Phys.",
    volume = "34",
    pages = "229--236",
    year = "1973"
}

@article{papapetrou-68,
     author = {Papapetrou, A. and Hamoui, A.},
     title = "{Surfaces caustiques d\'eg\'en\'er\'ees dans la solution de {Tolman.} {La} singularit\'e physique en relativit\'e g\'en\'erale}",
     journal = {Annales de l'institut Henri Poincar\'e. Section A, Physique Th\'eorique},
     pages = {343--364},
     publisher = {Gauthier-Villars},
     volume = {6},
     number = {4},
     year = {1967},
     url = {https://www.numdam.org/item/AIHPA_1967__6_4_343_0/}
}

@article{Tolman:1934za,
    author = "Tolman, Richard C.",
    title = "{Effect of imhomogeneity on cosmological models}",
    doi = "10.1073/pnas.20.3.169",
    journal = "Proc. Nat. Acad. Sci.",
    volume = "20",
    pages = "169--176",
    year = "1934"
}

@article{Bondi:1947fta,
    author = "Bondi, H.",
    title = "{Spherically symmetrical models in general relativity}",
    doi = "10.1093/mnras/107.5-6.410",
    journal = "Mon. Not. Roy. Astron. Soc.",
    volume = "107",
    pages = "410--425",
    year = "1947"
}

@article{Nolan:1999tw,
    author = "Nolan, Brien C.",
    title = "{Strengths of singularities in spherical symmetry}",
    eprint = "gr-qc/9902021",
    archivePrefix = "arXiv",
    doi = "10.1103/PhysRevD.60.024014",
    journal = "Phys. Rev. D",
    volume = "60",
    pages = "024014",
    year = "1999"
}

@book{leveque-92,
  author = {LeVeque, Randall J.},
  isbn = {978-3-7643-2723-1},
  pages = {1-214},
  publisher = {Birkhäuser},
  series = {Lectures in mathematics},
  title = {Numerical methods for conservation laws (2. ed.).},
  year = 1992
}

@Book{colombeau-92,
 Author = {Colombeau, Jean Fran{\c{c}}ois},
 Title = {Multiplication of distributions. {A} tool in mathematics, numerical engineering and theoretical physics},
 FSeries = {Lecture Notes in Mathematics},
 Series = {Lect. Notes Math.},
 ISSN = {0075-8434},
 Volume = {1532},
 ISBN = {3-540-56288-5},
 Year = {1992},
 Publisher = {Berlin: Springer-Verlag},
 Language = {English},
 DOI = {10.1007/BFb0088952},
 zbMATH = {108533},
 Zbl = {0815.35002}
}

@article{Israel:1967zz,
    author = "Israel, W.",
    title = "{Gravitational Collapse and Causality}",
    doi = "10.1103/PhysRev.153.1388",
    journal = "Phys. Rev.",
    volume = "153",
    pages = "1388--1393",
    year = "1967"
}

@book{Poisson:2009pwt,
    author = "Poisson, Eric",
    title = "{A Relativist's Toolkit: The Mathematics of Black-Hole Mechanics}",
    doi = "10.1017/CBO9780511606601",
    publisher = "Cambridge University Press",
    month = "12",
    year = "2009"
}

@article{lake-84,
  title = {The big bang in the Tolman models},
  author = {Lake, Kayll},
  journal = {Phys. Rev. D},
  volume = {29},
  issue = {4},
  pages = {771--772},
  numpages = {0},
  year = {1984},
  month = {Feb},
  publisher = {American Physical Society},
  doi = {10.1103/PhysRevD.29.771},
  url = {https://link.aps.org/doi/10.1103/PhysRevD.29.771}
}

@article{clarke-donnell,
    author =  {Clarke C.J.S. and O'Donnell N},
    title = {Dynamical extension through a space-time singularity},
    journal = {Rendiconti del
seminario matematico, Universit´a e Politecnico Torino},
    year = 1992
}

@ARTICLE{Frauendiener,
       author = {{Frauendiener}, J. and {Klein}, C.},
        title = "{On crossing dust shells.}",
      journal = {Journal of Mathematical Physics},
         year = 1995,
        month = jul,
       volume = {36},
       number = {7},
        pages = {3632-3643},
          doi = {10.1063/1.530987},
       adsurl = {https://ui.adsabs.harvard.edu/abs/1995JMP....36.3632F}
}

@book{lev02,
  author = {LeVeque, Randall J.},
  publisher = {Cambridge University Press },
  title = {Finite-Volume Methods for Hyperbolic Problems},
  year = 2002
}

@article{booth,
  title = {Closest safe approach to an accreting black hole},
  author = {Tippett, Benjamin K. and Booth, Ivan},
  journal = {Phys. Rev. D},
  volume = {90},
  issue = {8},
  pages = {084027},
  numpages = {14},
  year = {2014},
  publisher = {American Physical Society},
  doi = {10.1103/PhysRevD.90.084027},
  url = {https://link.aps.org/doi/10.1103/PhysRevD.90.084027}
}

@book{Dafermos:1315649,
      author        = "Dafermos, Constantine M",
      title         = "{Hyperbolic Conservation Laws in Continuum Physics; 3rd
                       ed.}",
      publisher     = "Springer",
      address       = "Dordrecht",
      series        = "Grundlehren der mathematischen Wissenschaften : a series
                       of comprehensive studies in mathematics",
      year          = "2010",
      url           = "https://cds.cern.ch/record/1315649",
      doi           = "10.1007/978-3-642-04048-1",
}

@article{Nolan:2004dm,
    author = "Nolan, B.",
    title = "{Singularity strengths and extendibility}",
    journal = "Conf. Proc. C",
    volume = "0405132",
    pages = "113--128",
    year = "2004"
}

@misc{Liu:2025fil,
      title={Quantum induced shock dynamics in gravitational collapse: insights from effective models and numerical frameworks}, 
      author={Hongguang Liu and Dongxue Qu},
      year={2025},
      eprint={2504.18462},
      archivePrefix={arXiv},
      primaryClass={gr-qc},
      url={https://arxiv.org/abs/2504.18462}, 
}
    
\end{document}